\documentclass[10pt]{article} 
\pdfoutput=1
\usepackage[utf8]{inputenc}
\usepackage[T1]{fontenc}
\usepackage[english]{babel}
\usepackage{geometry} 
\usepackage{authblk}
\usepackage{amsthm}
\usepackage{csquotes}
\usepackage{amsmath}
\usepackage{amsfonts}
\usepackage{amssymb}
\usepackage{bbold}
\usepackage[backend=biber, style=nature, sorting=none]{biblatex}
\usepackage[labelfont=bf]{caption}
\usepackage{changepage}
\usepackage{float}
\usepackage{graphicx}
\usepackage{mathrsfs}
\usepackage{mathtools}
\usepackage[arrowdel]{physics}
\usepackage{soul}
\usepackage{xcolor}
\usepackage{hyperref}
\hypersetup{colorlinks, allcolors=black}
\usepackage{siunitx}
\usepackage{fix-cm}

\DeclareUnicodeCharacter{03BC}{\ensuremath{\mu}}

\let\oldcite\cite
\renewcommand*\cite[1]{~\oldcite{#1}}

\title{\textbf{Controlling Mode-Locking in Liquid Light}} 
\author{Diego Piciocchi$^{1,\dagger,*}$, Francesco Fortuna$^{1,\dagger}$, Bahareh Marzban$^{1}$, \\ Lucius Miller$^{1}$, Ina Heckelmann$^{1}$, Laurenz Kulmer$^{2}$, \\ Juerg Leuthold$^{2}$, Jérôme Faist$^{1}$, Alexander Dikopoltsev$^{*}$}

\affil[1]{Institute for Quantum Electronics and Quantum Center, ETH Zürich, 8093 Zürich, Switzerland.}
\affil[2]{Institute for Electromagnetic Fields, ETH Zürich; 8092 Zürich, Switzerland. }
\affil[*]{\rm{Corresponding authors. Email: dpiciocchi@phys.ethz.ch, adikopoltsev@phys.ethz.ch}}
\affil[$\dagger$]{\rm{These authors contributed equally to this work.}}

\date{}

\begin{document}
  \maketitle

\begin{abstract}
    Mode-locking is the process by which the phases of many cavity modes organize collectively into a stationary state. In lasers, mode-locking originates from the interplay of gain, loss, dispersion and nonlinearities, most commonly resulting in the emission of short pulses. However, pulse formation is prevented when the gain recovery is fast enough to suppress amplitude fluctuations. Mode-locking then produces an extended state of nearly constant intensity, whose dynamics is governed by the phase. Yet, a general framework for designing the phase of such extended photonic states remains lacking. Here we introduce a hydrodynamic approach to designing extended photonic states, in which the phase dynamics are understood and controlled as a flow of liquid-like light. The intensity, clamped by fast gain, plays the role of a constant density, the instantaneous frequency that of a flow velocity, and the programmable radio-frequency modulation applied to the laser acts as a potential landscape that controls the flow. We derive analytical solutions connecting the applied modulation to the instantaneous frequency and hence to the spectrum emitted by the laser. We further provide experimental evidence for the liquid light shaping framework by measuring the frequency comb spectra emitted by a fast-gain semiconductor laser around \SI{1.6}{\micro\m} under the control of a programmed modulation, and comparing them to the ones predicted by the analytical solution. The hydrodynamic picture of mode-locking provides a predictive design principle for frequency combs, pointing toward a programmable synthesizer of light, and establishes fast-gain lasers as an accessible, electrically driven platform for nonequilibrium fluids of light.
\end{abstract}

\section{Introduction}

The locking of many oscillatory degrees of freedom into a common phase relationship is ubiquitous across physics, from synchronized pendulum clocks to injection-locked oscillators and arrays of Josephson junctions\cite{pikovskySynchronizationUniversalConcept2002, adlerStudyLockingPhenomena1946}. In each case, coupling among many oscillators selects a small set of collective coherent states from a much larger space of possibilities.

In photonics, this process is known as mode-locking. It requires a mechanism that couples cavity modes and then selects a steady state by constraining their amplitudes and phases. In active mode-locking, both mechanisms are effectively linear: a modulator transfers energy between neighboring modes, while dissipation selects the lowest-loss state, a pulse\cite{hausTheoryForcedMode1975,kuizengaFMAMMode1970}. However, both roles can be taken over by an optical nonlinearity, as in passively mode-locked lasers with saturable absorbers, Kerr lensing or additive-pulse mode locking\cite{mockerMODECOMPETITIONSELFLOCKING1965, demariaSELFMODELOCKINGLASERS1966, kellerSemiconductorSaturableAbsorber1996, spence60fsecPulseGeneration1991, markFemtosecondPulseGeneration1989}. Similarly, in coherently driven passive resonators, the Kerr nonlinearity couples the modes and maintains cavity and temporal solitons\cite{lugiatoSpatialDissipativeStructures1987, barlandCavitySolitonsPixels2002, herrTemporalSolitonsOptical2014, kippenbergDissipativeKerrSolitons2018}. The dynamics is then governed by master equations of the complex Ginzburg--Landau type, which balance gain, loss, dispersion and nonlinearity\cite{hausTheoryModeLocking1975, hausModelockingLasers2000, aransonWorldComplexGinzburgLandau2002, greluDissipativeSolitonsModelocked2012}. In all these cases, the selected state is localized in a pulse.

Mode-locked lasers with fast gain recovery, however, behave in a considerably different way. Gain that recovers faster than any mean-field timescale of the system cannot sustain strongly amplitude-modulated patterns\cite{khurginCoherentFrequencyCombs2014, singletonEvidenceLinearChirp2018}: gain suppresses intensity fluctuations and sets the intensity at a constant value. A finite recovery time, $T_1$, limits the response bandwidth, rendering gain saturation less effective for intensity variations faster than $1/T_1$\cite{burghoffCombsFastSlow2025}. The resulting state extends over the entire round-trip, while the dynamics is carried by the phase. When the modes are locked, the time-derivative of the phase, the instantaneous frequency, sweeps periodically. Mode-locking then requires the selection and stabilization of a phase profile under the constraint of nearly uniform intensity. Such frequency-modulated states are now established in quantum cascade\cite{hugiMidinfraredFrequencyComb2012, heckelmannQuantumWalkComb2023, piciocchiSpectralShapingFastgain2026}, interband cascade\cite{sterczewskiInterbandCascadeLaser2021}, quantum well\cite{daySimpleSinglesectionDiode2020, sterczewskiFrequencymodulatedDiodeLaser2020}, quantum-dash\cite{rosalesHighPerformanceMode2012} and quantum-dot lasers\cite{dongBroadbandQuantumdotFrequencymodulated2023}, as well as in optical parametric oscillators\cite{englebertTopologicalSolitonFrequency2025,stokowskiIntegratedFrequencymodulatedOptical2024}. Because the intensity does not drop to zero, the phase and the instantaneous frequency are defined at every point of the round trip, and the state is described by their local profiles along the cavity. Therefore, features related to local variations on the extended background can be studied, such as Nozaki--Bekki solitons\cite{opacakNozakiBekkiSolitons2024}, phase turbulence\cite{piccardoFrequencyCombsInduced2020}, and laminar--turbulent transitions\cite{turitsynaLaminarTurbulentTransition2013}. As these local properties also govern the robustness of the mode-locked states, they have been studied in the context of laser noise\cite{schneiderUltrafastNonHermitianSkin2025}, comb instabilities\cite{royLiquidCombsBroadband2025}, or preservation of collective oscillations\cite{dikopoltsevCollectiveQuenchDynamics2025}.

This description of extended laser states through their local properties connects naturally to the physics of quantum fluids of light\cite{carusottoQuantumFluidsLight2013}. In these systems, intensity variations and phase gradients play the roles of fluid density and velocity, while dispersion and nonlinear interactions govern their collective dynamics, providing a framework for phenomena such as sound waves\cite{utsunomiyaObservationBogoliubovExcitations2008, fontaineObservationBogoliubovDispersion2018}, superfluid flow\cite{amoSuperfluidityPolaritonsSemiconductor2009}, quantized vortices\cite{lagoudakisQuantizedVorticesExciton2008}, dark solitons\cite{amoPolaritonSuperfluidsReveal2011} and shock waves\cite{wan_dispersive_2007}. Lasers can be classified as incoherently pumped fluids of light, sustained by gain and loss, whose phase is selected by nonlinear dynamics rather than imposed by an external coherent drive\cite{woutersExcitationsNonequilibriumBoseEinstein2007, woutersSuperfluidityCriticalVelocities2010}. This connection suggests that the local properties of extended mode-locked states can also be understood as properties of a flowing fluid\cite{dikopoltsevCollectiveQuenchDynamics2025,piciocchiNoisedrivenLocalizationCollective2026}. For actively modulated fast-gain lasers, the challenge is to turn this interpretation into a predictive tool for stationary states of the flow, making the fluid description into a design algorithm for the laser field and its emitted spectrum.

\begin{figure}[ht]
    \centering
    \includegraphics[scale=1.0]{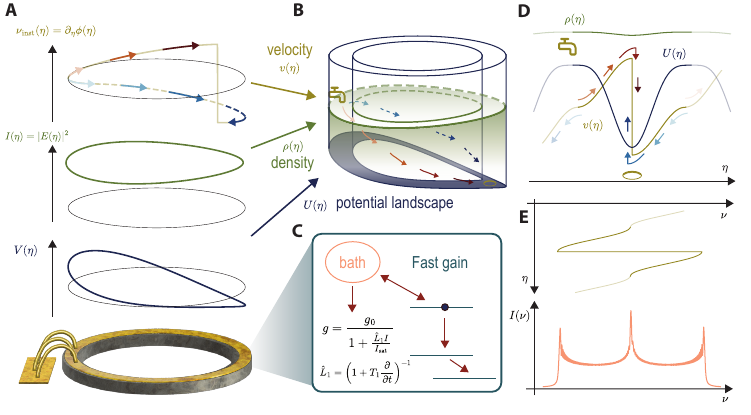}
    \caption{\textbf{Fast-gain lasers and liquid light.}
    (A) Unidirectional ring laser driven by a programmable radio-frequency waveform superimposed on the bias current. The resulting gain modulation is converted into a phase modulation through the laser's inherent nonlinearities, defining a periodic potential $U(\eta)$ in the co-rotating coordinate $\eta$. (B) Madelung mapping of the intracavity field: the optical intensity $\rho=|E|^2$ plays the role of a fluid density and the instantaneous frequency $v=\partial_\eta\varphi$ that of a flow velocity. (C) Fast gain (short $T_1$) suppresses fluctuations of the intensity, thereby clamping it to its steady-state value $\rho_0$ and confining the dynamics to the flow. (D) Stationary flow in the modulation landscape. The flow starts at rest from the maximum of the potential $U$, the splitting point, accelerates in both directions as the potential falls, and converges at an interface, where the instantaneous frequency jumps between the two branches. Distributed gain and loss are the non-conservative terms which sustain a non-uniform flow at constant density. (E) The comb spectrum follows from the flow, since the emitted amplitudes are the Fourier coefficients of $e^{i\varphi}$. }
    \label{fig:Fig1}
\end{figure}

Here, we establish a mapping between actively modulated fast-gain ring lasers (Fig.~\ref{fig:Fig1}A) and fluids of light, and exploit it to design their emitted spectra. Through a Madelung transformation\cite{madelungQuantentheorieHydrodynamischerForm1927}, the intensity of the field becomes the density of a fluid and its instantaneous frequency the flow velocity (Fig.~\ref{fig:Fig1}B--C). Since the fast gain enforces a constant intensity, the density is fixed and the light behaves as a one-dimensional liquid whose dynamics lies in the flow. Then, programmable radio-frequency modulation, which maps to a variable potential landscape, sets the velocity of this liquid  (Fig.~\ref{fig:Fig1} D). We find these velocities analytically and use the solution to design instantaneous frequency profiles, and their corresponding output spectra (Fig.~\ref{fig:Fig1} E). Unlike a mass-conserving one-dimensional fluid, whose flow must be uniform at constant density, an active photonic fluid can sustain a nonuniform flow at nearly constant density through gain and loss: it branches away from splitting points located at the potential maxima and meets with opposite velocities at interfaces where a discontinuity appears. We design peaked and flat-top comb spectra and provide experimental evidence for the analytical solution. Electrically programmable combs of this kind could find use in applications that require specific wavelengths with fast tunability, such as spectroscopy, where power can be concentrated on the absorption features, or optical communications, where power can be redistributed among channels.

\section{Liquid light in unidirectional fast-gain lasers}

We consider a circular-cavity laser of length $L$ that satisfies three conditions (Fig.~\ref{fig:Fig1}A). First, its gain recovers within a time $T_1$ shorter than other timescales of intracavity dynamics, so that gain saturation acts on the instantaneous intensity\cite{burghoffCombsFastSlow2025} and suppresses intensity fluctuations with frequencies smaller than $1/T_1$. In the model below, we take the instantaneous limit of the gain saturation shown in Fig.~\ref{fig:Fig1}C, which uses the low-pass filter form explicitly, and expand it to first order in intensity.
Second, the laser operates unidirectionally, so that no standing-wave grating forms and spatial hole burning does not cause mode mixing\cite{khurginCoherentFrequencyCombs2014, opacakFrequencyCombGeneration2021, seitnerBackscatteringInducedDissipativeSolitons2024}. 
Lastly, the coupling between modes is achieved through radio-frequency (RF) modulation of the current at the cavity resonance frequency $\Omega_R=Kc/n$, where $K=2\pi/L$. The asymmetry of the gain spectrum converts this modulation into phase modulation, which can be described through a direct linewidth enhancement factor (LEF) term, or equivalently by an effective Kerr-like nonlinearity\cite{opacakFrequencyCombGeneration2021, letsouHighpowerRingLaser2026, digiorgioSubharmonicInjectionQuantum2025}. 
In a frame co-moving with the drive, with coordinate $\eta=Kz\in[0,2\pi)$ and $z=x-(c/n)t$, the intracavity field obeys the complex Ginzburg--Landau equation\cite{dikopoltsevTheoryQuantumWalk2025}
\begin{equation}
    \partial_t E
    =\frac{\Gamma}{2}
    \left(1-\frac{|E|^2}{\rho_0}\right)E
    +\left( G_c+iD\right)\partial^2_\eta E - i U(\eta)E,
    \label{eq:master}
\end{equation}
where $\rho_0=I_s(1-\alpha_w/g_0)$ is the stationary intensity of the free-running laser. $g_0$, $\alpha_w$ and $I_s$ are the unsaturated gain factor, the loss and the saturation intensity. The factor $\Gamma=(g_0-\alpha_w)c/n$ is the rate at which the saturable gain restores the intensity towards $\rho_0$. $D$ is the group-velocity dispersion, $G_c$ the gain curvature, and $U(\eta)$ a real periodic potential proportional to the applied RF waveform\cite{heckelmannQuantumWalkComb2023, dikopoltsevCollectiveQuenchDynamics2025}. We consider $D>0$ and $G_c>0$; for dispersion $D$ with opposite sign, the roles of the potential maxima and minima in the following are exchanged.

In order to cast Eq.~\ref{eq:master} into a fluid equation, we use the Madelung transformation\cite{madelungQuantentheorieHydrodynamischerForm1927}, following the same approach used to derive the hydrodynamic formulation of the Schr\"odinger equation in its semiclassical limit. 
We first write the field as $E=\sqrt{\rho}\,e^{i\varphi}$ and define density as $\rho=|E|^2$, proportional to the intensity, and velocity as $v=\partial_\eta\varphi$, which is related to the instantaneous frequency through the free spectral range of the cavity, with $f_{ \mathrm{inst}}=( c/Ln)\cdot v$ (Fig.~\ref{fig:Fig1}B). We separate Eq.~\ref{eq:master} into intensity and phase dynamics: 
\begin{align}
    \partial_t\rho+2D\,\partial_\eta(\rho v)
    &= \Gamma\left(1-\frac{\rho}{\rho_0}\right)\rho
    + 2G_c\rho\left(Q-v^2\right),
    \label{eq:density}\\
    \partial_t\varphi
    &= -Dv^2-U+G_c\,\partial_\eta v
    + DQ + G_c\,v\,\partial_\eta\ln\rho .
    \label{eq:phase}
\end{align}
In these equations, the term $Q=\frac{\partial_\eta^2\sqrt\rho}{\sqrt\rho}$ is known as the quantum pressure\cite{carusottoQuantumFluidsLight2013}. With this reformulation of Eq.~\ref{eq:master}, all the terms describing the evolution of the laser's field have a role in the fluid. The term $Dv^2$ has a form of kinetic energy where $D$ plays the role of an inverse mass. The externally imposed modulation takes the role of a potential landscape $U(\eta)$ that results in a force $-\partial_\eta U$ for the flow. The saturable gain, $\Gamma\left(1-\frac{\rho}{\rho_0}\right)$, makes the fluid non-conservative and acts directly on the density, restoring it towards $\rho_0$ and suppressing density fluctuations. Gain curvature acts on both density and flow: it removes density where the field varies rapidly (through the term $-2G_c\rho v^2$), and smooths the flow through the term $G_c\partial_\eta v$, which is the effective viscosity of the fluid. It therefore provides a direct relaxation mechanism for the phase even when intensity variations are strongly suppressed by the gain saturation. The terms $Q$ and $\partial_\eta\ln\rho$ depend only on density gradients and vanish for uniform density.

Since fast gain saturation keeps the intensity close to $\rho_0$, we can write $\rho=\rho_0(1+\epsilon)$, where $\epsilon\ll 1 $. Imposing $\partial_t\rho=0$, we obtain the leading-order intensity correction $\epsilon\simeq-(2D\,\partial_\eta v+2G_c v^2)/\Gamma$ that is needed to sustain a non-uniform stationary flow. Where the flow diverges ($\partial_\eta v>0$), the intensity drops and the saturated gain increases, replenishing the outgoing light. Where the flow converges ($\partial_\eta v<0$), the opposite occurs. Gain curvature lowers the intensity further where $|v|$ is large, increasing the gain to compensate for the additional loss. Although a non-uniform intensity is needed to sustain the stationary flow, its feedback on the phase dynamics in Eq.~\ref{eq:phase} can be neglected. Gain saturation can also produce a Kerr-like phase shift through carrier-induced changes in the refractive index\cite{opacakFrequencyCombGeneration2021}. In quantum fluids of light, spatial gradients of such a density-dependent shift act as an interaction pressure\cite{carusottoQuantumFluidsLight2013}. Here, the small intensity variations give a negligible Kerr term compared to the RF potential depth, so we neglect this contribution.

Neglecting the impact of density variation and differentiating the equation for the phase, Eq.~\ref{eq:phase}, gives
\begin{equation}
    \partial_t v
    +2Dv\partial_\eta\!v+\partial_\eta U
    =G_c\partial_\eta^2v.
    \label{eq:flow}
\end{equation}
This is the forced viscous Burgers equation\cite{burgersMathematicalModel1948, hopfPartialDifferentialEquation1950}, which governs the evolution of the flow (that is, the instantaneous frequency). The potential drives the flow through the force $-\partial_\eta U$, while dispersion produces the nonlinear transport term $2Dv\,\partial_\eta v$, so that different parts of the profile propagate at different speeds and can steepen into shock fronts. Gain curvature acts as an effective viscosity, opposing this steepening and giving the fronts a finite width.

We now seek the stationary solution for the flow of this fluid. Since $v=\partial_\eta\varphi$, a stationary profile requires the phase differences between positions to remain constant, while the overall phase can evolve at a common rate. We therefore write $\varphi(\eta,t)=\Phi(\eta)-\omega t$, where $\omega$ is the common optical-frequency offset relative to the carrier. Substituting into Eq.~\ref{eq:flow} gives the stationary phase balance
\begin{equation}
Dv^2+U(\eta)-G_c\partial_\eta v=\omega,
\label{eq:bernoulli}
\end{equation}
which determines the flow together with the periodic boundary condition. Away from sharp transitions, the contribution of gain-curvature $G_c\partial_\eta v$ is small. Thus, Eq.~\ref{eq:bernoulli} reduces to $Dv^2+U\simeq\omega$, a Bernoulli balance for a pressureless fluid, in which the kinetic and potential contributions, $Dv^2$ and $U(\eta)$, sum to a constant along the flow. For a conventional incompressible fluid in a one-dimensional channel, mass conservation requires a spatially uniform velocity $\partial_\eta v =0$: a varying potential is balanced by pressure gradients. In our fluid, gain and loss restore the fluid density, allowing the flow to converge or diverge while the density remains nearly constant even without pressure.

To determine the stationary flow over the cavity, we impose the periodic boundary condition set by the ring geometry, $E(\eta+2\pi,t)=E(\eta,t)$. Since $v=\partial_\eta\Phi$, the integral $\oint v\,\dd\eta$ gives the net phase change around the ring, which must be an integer multiple of $2\pi$. We focus on the stationary state with the smallest loss from gain curvature, which penalizes large frequency excursions from the gain center. This state has zero net phase change, $\oint v\,\dd\eta=0$.
For a given $\omega$, the Bernoulli law gives two possible directions of flow, $v=\pm\sqrt{[\omega-U(\eta)]/D}$. To satisfy $\oint v\,\dd\eta=0$, the flow must include both signs and thus reverse from positive to negative and back over one round trip. Therefore, there must be two zeros for $v$ in the round-trip. Where positive flow meets negative flow, the two sides converge and form a narrow shock interface whose width is set by gain curvature\cite{burgersMathematicalModel1948}. In the opposite direction, the flow moves away on both sides and passes smoothly through zero at the splitting point $\eta_s$. As a consequence, the Bernoulli balance gives $\omega=U(\eta_s)$. For the flow to remain real around the ring, the argument of the square root must be nonnegative. Since $D>0$, this requires $\omega\geq U(\eta)$, so the splitting point lies at a global maximum of the potential, giving $\omega=U_{\max}$. When this maximum is unique, the flow leaves $\eta_s$ in opposite directions and converges at the interface $\eta_i$. Away from the interface, the stationary profile is
\begin{equation}
    v(\eta)\simeq
    \begin{cases}
        +\sqrt{[U_{\max}-U(\eta)]/D}, & \eta_s<\eta<\eta_i,\\
        -\sqrt{[U_{\max}-U(\eta)]/D}, & \eta_i<\eta<\eta_s+2\pi.
    \end{cases}
    \label{eq:branches}
\end{equation}
The periodic boundary condition for the selected state, $\oint v\,\dd\eta=0$, fixes the interface position $\eta_i$ by requiring the positive and negative contributions to cancel. Its position therefore depends on the potential over the entire round trip.

The stationary solution gives the instantaneous-frequency profile of the mode-locked state. Its stability relies on two complementary mechanisms that act on the intensity and phase. Gain saturation restores the intensity towards its steady-state value, while gain curvature stabilizes the phase profile by damping spatial phase perturbations. Gain curvature therefore does not contribute to the stationary balance away from the interfaces but is still essential for the nonlinear state stabilization.

\section{Spectral programming with liquid light}

The connection between modulation $U(\eta)$ and stationary flow $v$, and therefore the instantaneous frequency $f_{\mathrm{inst}}$, enables the programming of the spectrum emitted by the laser (Eq.~\ref{eq:branches}). By shaping the flow, the potential controls how the field sweeps through frequencies during each round trip. For nearly uniform intensity, the spectral envelope is determined by the time spent near each frequency, so that slowly varying portions of $v(\eta)$ favor spectral peaks, while a more uniform sweep tends to distribute power more evenly. This connection provides an intuitive guide to choosing the potential for a desired spectral shape. More formally, the resulting spectrum is calculated by integrating $v(\eta)$ to obtain the phase $\Phi(\eta)$ and evaluating the Fourier coefficients of $e^{i\Phi(\eta)}$.
\begin{figure}[htbp]
    \centering
    \includegraphics[scale=1]{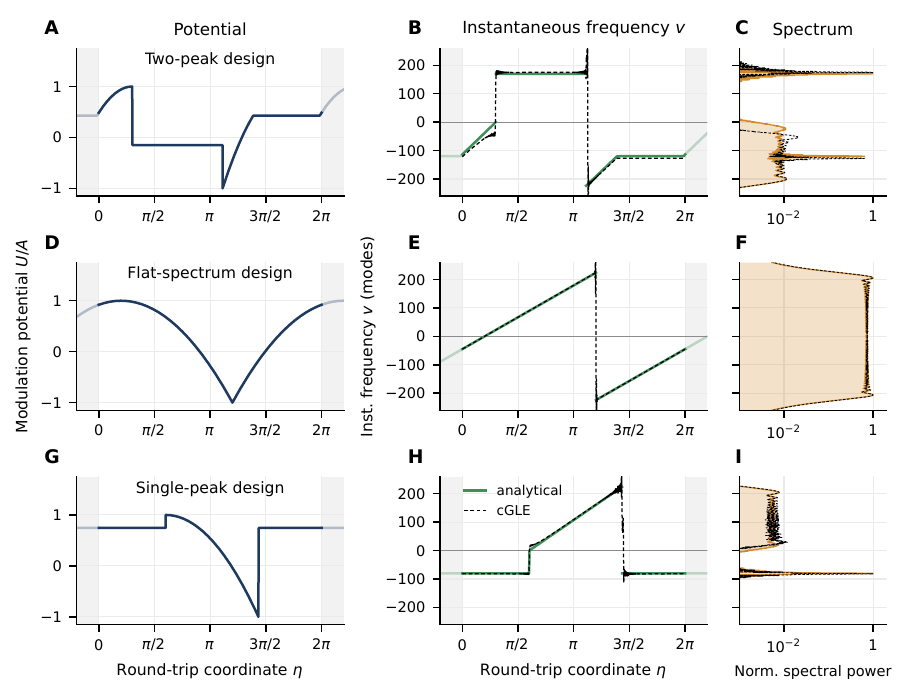}
    \caption{\textbf{Flexible shaping of liquid light.}
    Each row shows a designed modulation potential $U(\eta)$ (left), the resulting instantaneous-frequency profile $v(\eta)$ obtained from Eq.~\ref{eq:branches} and from Eq.~\ref{eq:master} (center), and the corresponding comb spectrum (right). (A--C) A modulation designed to produce a spectrum with two peaks of unequal amplitude. (D--F) A modulation producing a flat-top comb. (G--I) A modulation producing a comb with a single tunable peak. In each case the splitting point is located at the maximum of $U$ and the interface, visible as a jump in $v$, is fixed by the periodicity of the ring. Because the intensity is clamped, the spectrum is determined entirely by the flow.}
    \label{fig:Fig2}
\end{figure}

Figure~\ref{fig:Fig2} illustrates how this procedure produces different spectral shapes. The modulation in Fig.~\ref{fig:Fig2}A generates the frequency profile in Fig.~\ref{fig:Fig2}B, redistributing power into two peaks of unequal amplitude in Fig.~\ref{fig:Fig2}C.
Changing the potential reshapes the sweep to produce a flat-top comb in Figs.~\ref{fig:Fig2}D--F or a comb with a tunable peak in Figs.~\ref{fig:Fig2}G--I.
In each case, the spectrum follows from the same stationary flow construction, with the interface position fixed by phase periodicity.

\begin{figure}[h]
    \centering
    \includegraphics[scale=1.0]{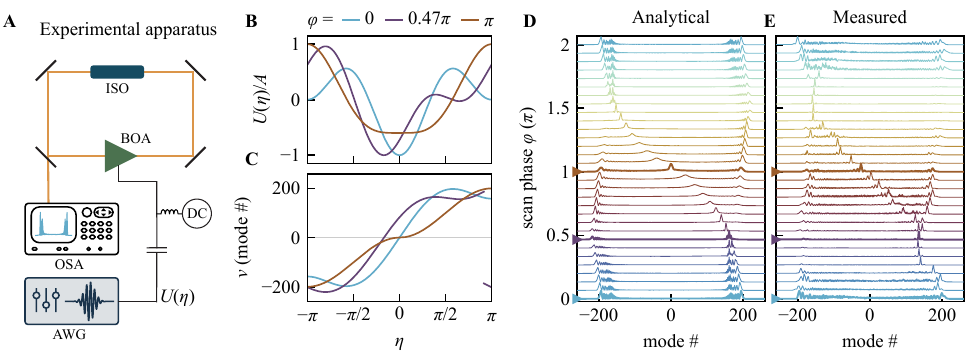}
    \caption{\textbf{Programmed flows and spectra.} (A) Experiment for programming the spectra of a fast gain laser. The free-space cavity contains the amplifier (BOA) and an optical isolator. The spectra are measured with an optical spectrum analyzer (OSA). The modulation is applied to the BOA using an arbitrary waveform generator (AWG). (B--C) Three representative programmed potentials $U(\eta)$ (top) and the resulting instantaneous-frequency profiles $v(\eta)$ (bottom), obtained from Eq.~\ref{eq:branches} for the injection of $f(\eta,\phi)$ with  phases $\varphi = 0$, $0.47\pi$, and $\pi$. (D) Analytical comb spectra as a function of the scan phase $\phi$, showing the mode occupation evolving as the phase is swept in the $0$ to $2\pi$ range. (E) Corresponding measured spectra, acquired by driving the laser with the same programmed waveforms over the same scan range.}
    \label{fig:Fig3}
\end{figure}

We validate this control scheme experimentally by applying resonant phase modulation to a unidirectional circular cavity laser with fast gain. The system is based on a free-space cavity incorporating a booster optical amplifier, as in the first demonstration of \SI{1.6}{\micro\m} quantum walk comb generation\cite{marzbanQuantumWalkComb2026}. We apply a \SI{300}{\mA} DC bias to the gain chip and modulate it with a programmed signal coming from an arbitrary waveform generator (AWG), resulting in a modulation depth $\dd J$ (Fig.~\ref{fig:Fig3}A). The modulation signal is constructed by combining two tones at $\Omega_R=2\pi\,\SI{838}{\mega\hertz}$ and $2\Omega_R$ with a relative phase $\phi$
\begin{equation}
U(\eta,\phi) = U_0 f(\eta, \phi), \qquad f(\eta,\phi) =-\frac{\cos\eta+M_2(\phi)\cos\left(2\eta+\phi\right)}{\mathcal N(\phi)},
\end{equation}
where $\mathcal N(\phi)$ is the maximum of the numerator and renormalizes the potential before it is rescaled by the amplitude. The amplitude of $M_2$, while programmed to be constant, evolves along the scan due to the response of the device, which we fitted as $M_2(\phi) = M_2(0) + \Delta M_2 \left[\frac{1+\cos\phi}{2}\right]$.

Conversion of the electrical signal to the potential entering Eq.~\ref{eq:branches} requires knowledge of the $D$ factor and rescaling of the amplitude. $D=\beta K^2/2$ is dictated by the group velocity dispersion $\beta=\SI{8320}{\fs^2/mm} \cdot c^3$. The amplitude $U_0$ is related to the modulation depth through $U_0=g_0\text{d}J\alpha$, where $g_0\approx\SI{0.19}{\cm^{-1}}$ is the unsaturated gain factor (obtained from the $\SI{30}{\dB}$ small-signal gain) and $\alpha=2$ is the amplitude-to-phase conversion factor. The current modulation depth was calibrated at $\dd J =7\%$ to match the experimental bandwidth of approximately 400 modes (that is, $\approx\SI{335}{\giga\hertz}$).

Three potential shapes are illustrated in Fig.~\ref{fig:Fig3}B, for three values of $\phi=0,0.47\pi,\pi$, and Fig.~\ref{fig:Fig3}C shows the corresponding instantaneous-frequency profiles predicted by Eq.~\ref{eq:branches}. Tuning $\phi$ between $0$ and $2\pi$ shapes the instantaneous frequency and the corresponding spectra, shown in Fig.~\ref{fig:Fig3}D. We compare these to the measured spectra of Fig.~\ref{fig:Fig3}E acquired over the same scan. The two maps show a compatible redistribution of power as the drive is varied, confirming that the flow construction predicts the qualitative comb shape and its control through modulation.

\section{Concluding Remarks}

We have formulated a hydrodynamic description of actively modulated fast-gain ring lasers and used it to design their output spectra. In this picture, optical intensity plays the role of fluid density and instantaneous frequency that of flow velocity. Fast gain saturation maintains a nearly uniform density, while gain curvature damps phase perturbations and stabilizes the flow selected by dispersion, modulation and cavity geometry. Gain saturation allows this liquid to sustain a spatially varying flow at nearly constant density by actively replenishing or removing light. This differs from a conventional incompressible fluid in a one-dimensional channel, where mass conservation requires a constant velocity.

This hydrodynamic picture turns the control of extended mode-locked states into a predictive design problem. An electrically programmed potential shapes the flow, determining the instantaneous frequency and the emitted comb spectrum. The measured spectral reshaping supports this connection between modulation and emission. The resulting design principle offers a route toward programmable frequency combs whose power distribution can be tailored to the needs of spectroscopy\cite{heckelmannBroadbandCombSpectroscopy2026}, ranging and optical communications.

Beyond spectral control, these results establish a connection between laser mode-locking and nonequilibrium fluid dynamics in a regime where density fluctuations are strongly suppressed. Fast-gain lasers thus offer an accessible, electrically driven setting for investigating collective phase dynamics, the stability of extended coherent states, and transitions to turbulence\cite{piccardoFrequencyCombsInduced2020, turitsynaLaminarTurbulentTransition2013}, with the same modulation providing a means to prepare and perturb the flow.

\section*{Acknowledgments}

D.P., F.F. and A.D. derived the analytical theory, and D.P. and F.F. performed the analytical calculations. \\
B.M. and L.M. assembled and initially characterized the device used in the experiment.\\
D.P. and A.D. performed the measurements with the assistance of L.K., using equipment provided by L.K. and J.L. \\
I.H. and A.D. conceived the idea of controlling the comb through modulation and A.D. proposed the idea of the liquid-light mapping. \\
D.P. wrote the manuscript with the assistance of A.D. and F.F. \\
A.D., J.L. and J.F. acquired and administered funding.\\

I.H. and A.D. thank Bálint Kaszás and Jeff Maki for insightful discussions on liquid light. A.D. and D.P. acknowledge Prof. Oded Zilberberg and Prof. Iacopo Carusotto for fruitful discussions on the topic of liquid light. 

\bigskip 
Data underlying this work are available from the authors upon request.

\printbibliography[title={References}]

\end{document}